\documentclass[prb,showpacs,twocolumn]{revtex4}

\usepackage{amsmath,amssymb,epsfig,epsf}

\usepackage{color}

\newcommand \modified[1]{\textcolor{black}{#1}}

\begin{document}

\title
{Consistent LDA$^\prime$+DMFT approach to electronic structure of transition 
metal oxides: charge transfer insulators and correlated metals.}
\author{$^1$I.A. Nekrasov, $^1$N.S. Pavlov, $^{1,2}$M.V. Sadovskii}

\affiliation
{$^1$Institute for Electrophysics, Russian Academy of Sciences, Ural Branch,\\ Amundsen str. 106, Ekaterinburg, 620016, Russia\\ 
$^2$Institute for Metal Physics, Russian Academy of Sciences, Ural Branch, \\  S. Kovalevskaya str. 18, Ekaterinburg, 620990, Russia}

\begin{abstract}

We discuss the recently proposed LDA$^\prime$+DMFT approach 
\modified{providing consistent parameter free treatment of}
the so called double counting problem arising within the 
LDA+DMFT hybrid computational method for realistic strongly correlated materials. 
In this approach the local exchange-correlation portion of electron-electron 
interaction is excluded from self consistent LDA calculations
for strongly correlated electronic shells, e.g. $d$-states of transition 
metal compounds. Then the corresponding double counting term in 
LDA$^\prime$+DMFT Hamiltonian is \modified{consistently} set in the local Hartree 
(fully localized limit - FLL) form \modified{of the Hubbard model interaction term}. 
We present the results of extensive LDA$^\prime$+DMFT calculations of
densities of states, spectral densities and optical conductivity
for most typical representatives of two wide classes of strongly correlated 
systems in paramagnetic phase:  charge transfer insulators (MnO, CoO and NiO) 
and strongly correlated metals (SrVO$_3$ and Sr$_2$RuO$_4$).
It is shown that for NiO and CoO systems  LDA$^\prime$+DMFT
qualitatively improves the conventional LDA+DMFT results with FLL type of 
double counting, where CoO and NiO were obtained to be metals. We also include
in our calculations transition metal 4$s$-states located near the Fermi level 
missed in previous LDA+DMFT studies of these monooxides. General agreement with
optical and X-ray experiments is obtained. For strongly correlated metals 
LDA$^\prime$+DMFT results agree well with earlier LDA+DMFT calculations and 
existing experiments. However, in general LDA$^\prime$+DMFT results give 
better quantitative agreement with experimental data for band gap sizes and 
oxygen states positions, as compared to the conventional LDA+DMFT.

\end{abstract}


\maketitle

\section{Introduction}

During last decade the LDA+DMFT method (local density approximation + dynamical 
mean-field theory) became probably the most powerful tool to calculate electronic 
structure of real strongly correlated materials 
[\onlinecite{poter97,LDADMFT1,Nekrasov00, psik, LDADMFT,IzAn}].
Typically this approach consists of two computation steps.
First, LDA calculations are exploited to obtain the non-interacting Hamiltonian 
$\hat{H}^{\rm LDA}$ which describes, rather accurately, the kinetic energy (and 
to some extent takes into account electronic interactions). At the second step 
the local Coulomb (Hubbard) interaction $\hat{H}^{\rm Hub}$ is introduced
into the lattice problem defined by $\hat{H}^{\rm LDA}$ for those electronic 
shells which are supposed to be strongly correlated. Thus obtained generalized 
Hubbard model is solved numerically using DMFT. Some attempts to organize a
feedback from DMFT step back to LDA calculations to achieve fully 
self-consistent LDA+DMFT are also known and may be important for some physical 
problems [\onlinecite{scLDADMFT}].

The double counting problem arises in the standard LDA+DMFT, because some 
portion of local electron-electron interaction for correlated shells is actually
accounted for within $\hat{H}^{\rm LDA}$. To avoid this double counting it is
necessary to subtract a certain correction term $\hat{H}^{DC}$ from 
$\hat{H}^{\rm LDA}$. Then the formal LDA+DMFT Hamiltonian is written as:
\begin{eqnarray} 
\hat{H} &=& \hat{H}^{\rm LDA} + \hat{H}^{\rm Hub} - \hat{H}^{DC}.
\label{HLDADMFT}
\end{eqnarray}
In orbital space $\hat{H}^{DC}$ is the diagonal matrix with non zero and equal
matrix elements \modified{$E_{dc}$} for these atomic shells assumed to be strongly 
(e.g. $d$ or $f$ shells or their subshells).
This becomes more transparent if we consider the corresponding Green's function for the 
Hubbard model:
\begin{equation}
\hat{G}_{ij}(\textbf{k}E)=[(E - \mu) \hat{I}-H^{\rm LDA}_{ij}(\textbf{k}) - (\Sigma(\textbf{k}E) - E_{dc}) \delta_{id}\delta_{jd}]^{-1},
\label{Dyson_for_DMFT}
\end{equation}
where $\hat{I}$ is the unity matrix in the orbital space, $\mu$ is the chemical 
potential and $\Sigma(\textbf{k}E)$ is the self-energy corresponding to local 
Coulomb (Hubbard) interaction, $[...]^{-1}$ denotes matrix 
inversion, while index $d$ denotes correlated states for which Coulomb 
(Hubbard) interaction is taken into account.

From Eq.~\eqref{Dyson_for_DMFT} one can see that in case of $\hat{H}^{\rm LDA}$ 
containing only the contribution of interacting $d$ - orbitals, $E_{dc}$ reduces 
to trivial renormalization of the chemical potential $\mu$. Then, strictly 
speaking, there is no double counting problem at all. Because of this many of 
early works (listed e.g. in reviews [\onlinecite{LDADMFT1, psik, LDADMFT,IzAn}], 
except probably the first paper on LDA+DMFT~[\onlinecite{poter97}] and few 
others) just dropped the double counting correction term. Only after the 
LDA+DMFT community started the active studies of multiband $\hat{H}^{\rm LDA}$ 
Hamiltonians with both correlated and non correlated states included, the
problem of correct implementation of $\hat{H}^{DC}$ became important.
Now there are dozens of works devoted to multiband LDA+DMFT 
studies. Important classes of materials investigated can be listed as: 
\begin{enumerate}
\item Transition metal oxides
(LaTiO$_3$, (Sr,Ca)VO$_3$, V$_2$O$_3$, VO$_2$, CrO$_2$, LaMnO$_3$, NiO, MnO, CoO, FeO,
LaCoO$_3$, TiOCl, Tl$_2$Mn$_2$O$_7$, LaNiO$_3$, (Ca,Sr)$_2$RuO$_4$, Na$_{0.3}$CoO$_2$);
\item Elemental transition metals and non-oxide transition metal compounds
(Cr, Mn, Fe, Ni, Co, multilayers (CrAs)/(GaAs), NiMnSb, Co$_2$MnSi, CrAs, VAs, ErAs,
Ni(S,Se)$_2$, KCuF$_3$);
\item Elemental $f$-electron materials and their compounds
(Ce, Pu, Am, Ce$_2$O$_3$, Pu$_2$O$_3$, USe, UTe, PuSe, PuTe, PuCoGa$_5$, URu$_2$Si$_2$,
CeIrIn$_5$, CeCoIn$_5$, CeRhIn$_5$);
\item Nano materials (Ni-Cu nano contacts and nano electrodes);
\item High temperature copper superconductors ((Sr,La)$_2$CuO$_4$, (Pr,Ce)$_2$CuO$_4$,
Bi$_2$Ca$_2$SrCuO$_8$ etc.);
\item Superconducting iron pnictides (LaFeAsO, CeFeAsP, LiFeAs, BaFe$_2$As$_2$, etc.).
\end{enumerate}
These systems show a large variety of physical effects. Among them there are 
strongly correlated metals, Mott and charge transfer insulators, ferromagnets 
and antiferromagnets, superconductors, etc. However, up to now there is no 
universal and unambiguous expression for $\hat{H}^{DC}$, and different
formulations are used for different classes of materials.

In this paper we present the results of extensive application 
of our recently proposed LDA$^\prime$+DMFT [\onlinecite{cLDADMFT}] approach to
charge transfer insulators MnO, CoO and NiO and
strongly correlated metals SrVO$_3$ and Sr$_2$RuO$_4$, confronted to conventional LDA+DMFT results and some experiments.
The manuscript has following structure. In Sec.~II we present an overview of 
different definitions of the $\hat{H}^{DC}$. The novel consistent 
LDA$^\prime$+DMFT method is described in Sec.~III. LDA and 
LDA$^\prime$ band structures, total and partial densities of states, spectral density maps and 
optical conductivity LDA$^\prime$+DMFT results for prototype charge transfer 
insulators MnO, NiO and CoO are presented in Sec.~IV and compared with the
results of conventional LDA+DMFT. These results are further compared with 
experimental data on X-ray spectroscopy and optical conductivity. In Sec.~V we discuss LDA and 
LDA$^\prime$ band structures for correlated metallic systems prototypes 
SrVO$_3$ and Sr$_2$RuO$_4$ are presented. Then LDA+DMFT and LDA$^\prime$+DMFT 
results are compared with each other and with experimental photoemission
and absorption spectra. Finally we end up with the Conclusion (Sec.~VI).

\section{Review of different formulations for $\hat{H}^{DC}$}

To derive an expression for $\hat{H}^{DC}$ let us examine $\hat{H}^{\rm LDA}$ 
and  $\hat{H}^{\rm Hub}$ terms Eq.~\eqref{HLDADMFT}.
LDA part of the Hamiltonian \eqref{HLDADMFT} is given by:
\begin{eqnarray}
\hat{H}_{{\rm LDA}} &=& -\frac{\hbar ^{2}}{2m_{e}}\Delta +V_{{\rm ion}}({\bf r})
+\int d^3{r^{\prime }}\,\rho ({\bf r^\prime })V_{ee}({\bf r}\!-\!{\bf r^\prime})\nonumber \\
&+&\frac{\delta E_{\rm xc}^{\rm LDA}(\rho )}{\delta \rho({\bf r})}
\label{HLDA0},
\end{eqnarray}%
where $\Delta $ is the Laplace operator, $m_{e}$ the electron mass, 
$e$ the electron charge, and
\begin{eqnarray}
V_{{\rm ion}} ({\bf r})=-e^2\sum_i \frac{Z_i}{|{\bf r}-{\bf R_i}|},
&\;\; & V_{\rm ee}({\bf r}\!-\!{\bf r'})=\frac{e^2}{2}
\sum_{{\bf r} \neq {\bf r'}} \frac{1}{|{\bf r}-{\bf r'}|}\nonumber\\
\end{eqnarray}
denote the one-particle potential due to all ions $i$
with charges $eZ_{i}$ at given positions ${\bf R_{i}}$, and the
electron-electron interaction, respectively.

The $E_{\rm xc}^{\rm LDA}(\rho ({\bf r}))$ in \eqref{HLDA0} is some function of
local charge density, which approximates the true exchange correlation
functional $E_{\rm xc}[\rho ]$ of density functional theory within 
local density approximation [\onlinecite{JonesGunn}].
The explicit expression for $E_{\rm xc}^{\rm LDA}(\rho ({\bf r}))$
is usually derived from perturbation theory [\onlinecite{jellium}]
or numerical simulations [\onlinecite{jellium2}] of the ``jellium'' model with 
$V_{\rm ion}({\bf r})={\rm const}$.
To obtain the value of local charge density one should
choose some basis set of one-particle wave functions $\varphi _{i}$
(e.g. to do practical calculations and explicitly express matrix elements 
of the Hamiltonian \eqref{HLDA0}), so that $\rho({\bf r})$ is written as:
\begin{equation}
\rho ({\bf r})=\sum_{i=1}^{N}|\varphi _{i}({\bf r})|^{2}.
\label{rhophi}
\end{equation}%

\modified{Hubbard-like (local) interaction term including direct Coulomb interaction and
exchange Coulomb interaction contributions in the density-density form is written as:}
\begin{eqnarray}
\hat{H}^{Hub} &=&U\sum_{m}\sum_{i}\hat{n}_{im\uparrow}\hat{n}_{im\downarrow }\label{H} \nonumber \\
&+&\;\sum_i\sum_{m\neq m'}\sum_{\sigma \sigma'}\;
(U'-\delta_{\sigma \sigma'}J)\;
\hat{n}_{im\sigma}\hat{n}_{im'\sigma'}.
\label{Hamiltonian}
\end{eqnarray}
\modified{Here, the index $i$ enumerates lattice sites, $m$ denotes orbitals, and
$\sigma$ the spin. The $U$ represents local intra-orbital Coulomb 
repulsion and $J$ -- $z$-component of Hund's rule coupling between the strongly correlated 
electrons (e.g. $d$-states, enumerated by $i=i_{d}$ and $l=l_{d}$).
Rotational invariance then fixes the local inter-orbital Coulomb repulsion
$U^\prime=U-2J$~\cite{Zoelfl00}.
The values of $U$ and $J$ are obtained usually from constrained
LDA procedure [\onlinecite{Gunnarsson}].
}
One can get numerically exact solution of the Hubbard Hamiltonian
(simplified kinetic term plus $\hat{H}^{\rm Hub}$ term) within DMFT 
approximation.

Hamiltonian $\hat{H}_{\rm LDA}$ contains local electron-electron correlations
through the exchange correlation energy (taken in the form valid for uniform 
electronic gas) and density-density contribution of the Hartree term. In its
turn, DMFT provides the numerical solution of the Hubbard model (exact in
infinite dimensions). Thus it is clear that before plugging $\hat{H}_{\rm LDA}$ 
into DMFT lattice problem \eqref{Dyson_for_DMFT}, one must subtract certain 
double counting correction term $\hat{H}^{DC}$ from $\hat{H}_{\rm LDA}$.
The double counting {\it problem} arises because there is no explicit
microscopic or diagrammatic relation between the model (Hubbard like) 
Hamiltonian approach and LDA. \modified{There} is apparently no possibility to give 
a rigorous expression for $\hat{H}^{DC}$ in terms of $U$, $J$ and 
$\rho$. \modified{Thus,} several $ad~hoc$ expressions for $\hat{H}^{DC}$ and approaches to 
\modified{treat} the double counting problem exist in the current literature. Below we 
briefly discuss some of these derivations.

\modified{Perhaps for the first time problem of double counting appeared within an attempt 
to merge LDA and the Hubbard model within the LDA+U method [\onlinecite{LDAU}],
where was initially {\em postulated} the so called ``around mean-field'' (AMF) definition of $\hat{H}^{DC}$.
This definition comes from an assumption that LDA is a kind of ``mean-field'' solution of the
Hubbard-like problem Eq.~\eqref{Hamiltonian}.
Later on the definition of Ref.~[\onlinecite{LDAU}] was generalized for spin dependent (LSDA)
case (and even more general -- with matrix form of Coulomb interaction).
After this spin dependent generalization corresponding AMF expression can be given as:}
\begin{equation}
\hat{H}_{AMF}^{DC}=\frac{1}{2}U\sum_{\sigma}n_{d\sigma}(n_d-n^0_\sigma)-
                   \frac{1}{2}J\sum_{\sigma}n_{d\sigma}(n_{d\sigma}-n^0_\sigma)
\label{dc_amf}
\end{equation}
\modified{with the average occupancies $n^{0}=\frac{1}{2(2l+1)}\sum_{m,\sigma}n_{m\sigma}$,
$n^{0}_\sigma=\frac{1}{(2l+1)}\sum_{m}n_{m\sigma}$ and total number of
electrons on interacting orbitals (per spin projection)
$n_{d\sigma}=\sum_{m}n_{il_{d}m\sigma}=\sum_{m}\langle \hat{n}_{il_{d}m\sigma}\rangle $
and $n_d=\sum_\sigma n_{d\sigma}$.
originally supposed to be found from LDA calculations.}
The drawback of AMF is the equal occupancy of all orbitals which 
is not correct even for weakly correlated systems because e.g. of crystal field 
splitting. However, a couple of the modern LDA+DMFT works reported the
reasonable results with AMF-like double counting correction term. Apparently,
the AMF double counting correction works rather well for moderately
correlated metallic systems. Some modifications of \eqref{dc_amf} were given in 
Refs.~[\onlinecite{Kunes}] and applied to LDA+DMFT calculations for charge 
transfer insulators.

Later on the fully localized (or atomic) limit (FLL) expression for 
$\hat{H}^{DC}$ was introduced in Refs.~[\onlinecite{fll1,fll2}] (with first 
application to LDA+DMFT calculations in Ref.~[\onlinecite{poter97}]):
\begin{equation}
\hat{H}_{FLL}^{DC}=\frac{1}{2}U n_{d}(n_{d}-1)-
\frac{1}{2}{J}\sum_\sigma n_{d\sigma}(n_{d\sigma}-1).
\label{ELDAU}
\end{equation}%
%
\modified{
The Eq.~\eqref{ELDAU} actually represents the  Hartree decoupling 
of the Hubbard model interaction term (\ref{Hamiltonian}) ---
decoupling of the density-density term $\hat{n}_i\hat{n}_j$
and not full four operator term $\hat{c}^\dagger_i\hat{c}^\dagger_j\hat{c}_o\hat{c}_l$.
Thus strictly speaking in Eq.~\eqref{ELDAU} there is no
Fock type of contribution since Hund exchange is presented in Eq.~(\ref{Hamiltonian})
in the density-density form,
although Hund coupling value $J$ has ``exchange nature''.
Quite often it is misinterpreted as due to the ``true'' Hartree-Fock decoupling of
$\hat{c}^\dagger_i\hat{c}^\dagger_j\hat{c}_o\hat{c}_l$ term.
}

The FLL expression in the context of LDA+DMFT calculations was used in the
majority of modern works. It works reasonably good for both metallic and
insulating strongly correlated materials.
Recently some modifications of FLL were proposed in 
Refs.~[\onlinecite{Zhu,AnisFe}]. Typically these modifications are used to 
quantitative improvements of LDA+DMFT results for particular compounds.
Some kind of AMF and FLL ``hybrid scheme'' was used in 
Ref.~[\onlinecite{AnisaFe}] for $\alpha-Fe$.

Alternative way to derive or guess the $\hat{H}^{DC}$ term 
is to express it through the characteristics of intrinsic single DMFT impurity 
problem, such as impurity self-energy $\Sigma^{imp}_{mm^\prime}$ or impurity 
Green's function $G^{imp}_{mm^\prime}$. A popular way is to define double 
counting energy as a static part of the impurity self-energy [\onlinecite{KLK}]:
\begin{equation}
E_{dc}=\frac{1}{2}\mathrm{Tr}_{\sigma}(\Sigma^{imp}_{\sigma}(0)).
\end{equation}%
Some of LDA+DMFT papers used this definition in calculations of metallic 
magnetic and non-magnetic systems. From the very beginning this type of double 
counting correction was also exploited within the GW+DMFT approach 
[\onlinecite{gwdmft}].

Hartree energy can be determined from LDA+DMFT self-energy as
its real part in the high frequency limit value. In Ref.~[\onlinecite{Karolak}] 
it was proposed to use thus defined Hartree energy as a double counting 
correction, using the constraint
\begin{equation}
\mathrm{Re}\mathrm{Tr}(\Sigma^{imp}_{mm^\prime}(i\omega_N))\overset{}{=}0,
\end{equation}%
where $\omega_N$ is the highest Matsubara frequency (used in calculations).
Physically similar definition of double counting term $E_{dc}=\Sigma({\omega\to\infty})$ was successfully 
applied to metallic ferromagnetic SrCoO$_3$[\onlinecite{sinfty}].

For metallic systems it was suggested to fix the double counting correction
by equating the number of particles of non-interacting problem and impurity 
problem as expressed via corresponding Green's function  [\onlinecite{gfdc}]:
\begin{equation}
\mathrm{Tr}~G^{imp}_{mm^\prime}(\beta)\overset{}{=}\mathrm{Tr}~G^{0,loc}_{mm^\prime}(\beta),
\label{trace}
\end{equation}
\modified{where $G^{0,loc}_{mm^\prime}$ is local non interacting Green function.}
Some of LDA+DMFT works treated double counting energy $E_{dc}$ as a free 
parameter. The authors of Ref.~[\onlinecite{Karolak}] found that most of 
described $\hat{H}^{DC}$ terms proposed in the literature are not completely 
satisfactory in the case of charge transfer insulator NiO and proposed 
a {\it numerical} way to define the necessary double counting correction.

Another possible solution of the double counting problem is
to perform Hartree+DMFT or Hartree-Fock+DMFT calculations [\onlinecite{Held}].
While performing Hartree-Fock band structure calculations for real materials
we do know exactly what portion of interaction is included.
Since diagrammatic expression for Hartree or Hartree-Fock terms are well known,
one can calculate them directly and get double counting correction energy
explicitly. However, up to now we are unaware of any Hartree+DMFT
or Hartree-Fock+DMFT calculations for real materials.

Completely independent branch of {\it ab initio} DMFT calculations is GW+DMFT 
method, which uses instead of density functional theory the so called chain of 
Hedin equations truncated in a simplest manner by the neglect of vertex 
corrections (for review see Ref.~[\onlinecite{gwdmft,HeldGW})].
Because of purely diagrammatic nature of GW there is a natural way to calculate 
the local part of corresponding Hartree contribution, which can be used as the
double counting correction term for GW+DMFT.~[\onlinecite{HeldGW}].

\section{Consistent LDA$^{\prime}$+DMFT approach}
\label{method}

Recently we proposed the LDA$^\prime$+DMFT approach, 
which defines consistent parameter free way
to avoid the double counting problem [\onlinecite{cLDADMFT}]. The main
idea is to exclude explicitly exchange-correlation energy from 
self-consistent LDA calculations only for correlated bands.
As described above main obstacle to express double counting term
exactly is exchange correlation $E_{\rm xc}^{\rm LDA}(\rho ({\bf r}))$
portion of interaction within LDA. So it seems somehow inconsistent to use it to 
describe correlation effects in narrow (strongly correlated) bands from the very
beginning, as these should be treated via more elaborate schemes like DMFT.  
To overcome this difficulty for these states, we propose 
to redefine charge density (\ref{rhophi}) in $E_{\rm xc}^{LDA}$ as follows:
\begin{equation}
\rho^\prime ({\bf r})=\sum_{i\neq i_d}|\varphi _{i}({\bf r})|^{2}
\label{rhophi1}
\end{equation}
{\em excluding the contribution of the density of strongly correlated electrons}.

In principle $E_{\rm xc}^{LDA}$ is not an additive function of charge density.
Thus splitting of charge density into two parts may lead to some loss of hybridization
between correlated and uncorrelated states. However as we show below this approximation
is rather good. Later on we will see that LDA$^\prime$ bands practically
do not change their shape with respect to LDA ones for all considered systems.
That tells us that ``hybridization'' is almost not affected by LDA$^\prime$. The main effect is
increase of splitting between oxygen 2p and metal 3d states. It comes from more repulsive potential
appearing in the LDA$^\prime$ case since part of exchange correlation energy is excluded there.

Then this redefined $\rho^\prime ({\bf r})$ (\ref{rhophi1}) is used to obtain 
$E_{\rm xc}^{\rm LDA}$ and perform the self-consistent LDA$^\prime$ band structure 
calculations for correlated bands. 
This procedure leaves out of interaction for correlated states on the LDA$^\prime$ stage
just the Hartree contribution (\ref{HLDA0}).
Thus, double counting correction term should be consistently taken
in the form of the Hartree like term, given by  Eq.~(\ref{ELDAU}).
This $H^{DC}_{FLL}$ definition also does not have any free parameters.
Actually, our approach is in precise correspondence with the standard definition of
correlations, as interaction corrections ``above'' the Hartree-Fock.
At the same time all other 
states (not counted as strongly correlated) are to be treated with the full 
power of DFT/LDA and {\em full} $\rho$ in $E_{\rm xc}^{LDA}$.

Despite the fact that the LDA$^\prime$+DMFT method is apparently most consistent 
with the use of FLL type of double counting, in principle all mentioned above definitions of $H^{DC}$ can also be exploited
within LDA$^\prime$+DMFT.
Also there is another ``degree of 
freedom left'' --  the occupancy $n_d$, used in FLL equation, can be obtained 
either from LDA or LDA$^\prime$ results, or it can be calculated
self consistently during the DMFT loop. We used all these variants in our
calculations for different compounds presented below. 
Corresponding values of $E_{dc}$ listed in Table~\ref{tab1}.
Notations are: FLL(SC) for self consistently calculated $n_d$ and
for $n_d$ from LDA or LDA$^\prime$ -- FLL(LDA).
In general FLL(SC) and FLL(LDA) do not differ very much from each other, except 
for the case of CoO (see below). However, FLL(SC) gives slightly better 
agreement with experiments. Most Figures presented below are plotted for 
the FLL(SC) case. We observed that FLL(SC) calculations require more computational 
time than FLL(LDA).

Thereby our consistent LDA$^{\prime}$+DMFT approach 
is a kind of compromise between Hartree-Fock and DFT/LDA 
starting points to be followed by DMFT calculations.
It was demonstrated in Ref.~[\onlinecite{cLDADMFT}]
that this LDA$^\prime$+DMFT method works perfectly for insulating NiO system,
directly producing charge transfer insulator solution, while conventional 
LDA+DMFT (with FLL) gives metallic solution (cf. Ref.~[\onlinecite{Karolak}]).

\section{Charge transfer insulators}
\label{cti}

\subsection{LDA and LDA$^\prime$ band structures}

\begin{figure}[!ht]
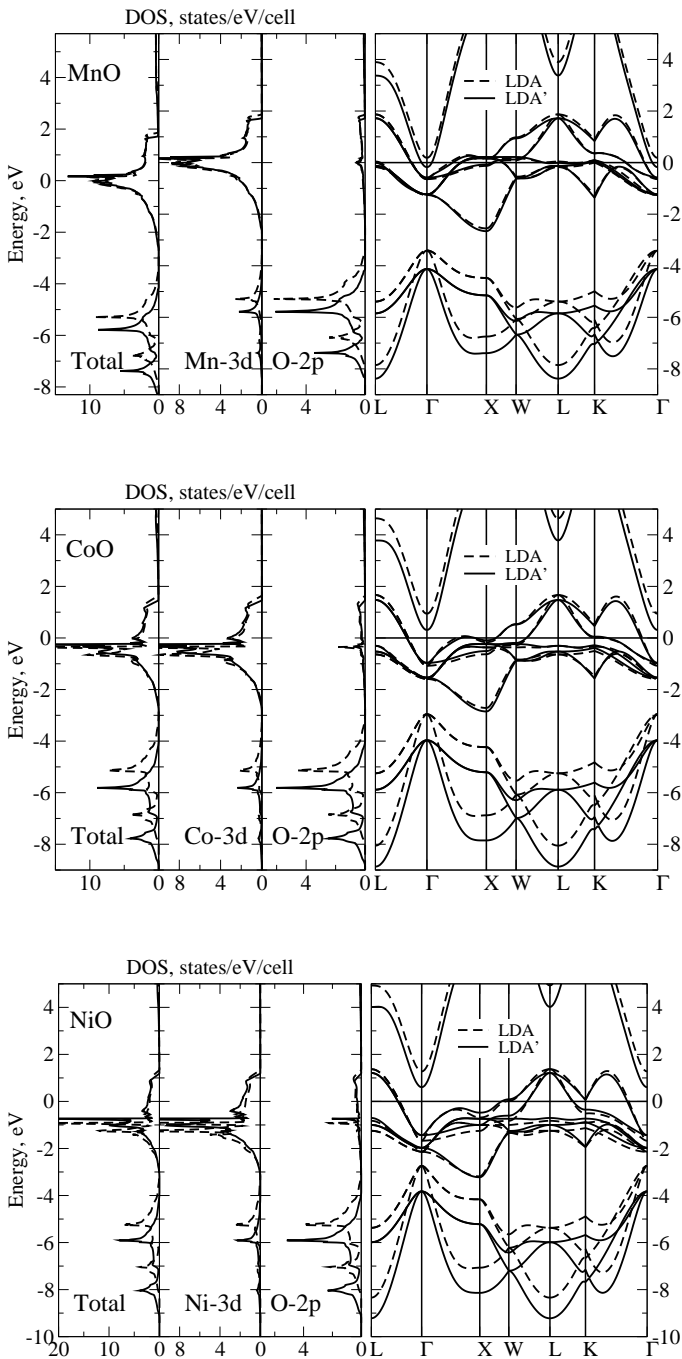

\includegraphics[width=.5\textwidth]{MnO_LDA_bands_and_DOS_comp.eps}\newline
\vspace{.5cm}
\includegraphics[width=.5\textwidth]{CoO_LDA_bands_and_DOS_comp.eps}\newline
\vspace{.5cm}
\includegraphics[width=.5\textwidth]{NiO_LDA_bands_and_DOS_comp.eps}
\caption{LDA (dashed lines) and LDA$^\prime$ (solid lines) densities of states 
(DOS) and band dispersions for MnO (upper row),
CoO (middle row) and NiO (lower row). Fermi level is zero.
}
\label{fig6}
\end{figure}

Typical examples of charge transfer insulator (CTI) materials are transition metal monoxides MnO, CoO and 
NiO. These oxides have rock salt crystal structure with
lattice parameters a=4.426\AA, 4.2615\AA~and 4.1768\AA~correspondingly.
To obtain LDA and LDA$^\prime$ band structures for  MnO, CoO and NiO
the linearized muffin-tin orbitals (LMTO) basis set [\onlinecite{LMTO}] was used.
In the corresponding program package TB-LMTO v.47 the $E_{\rm xc}^{\rm LDA}$
was taken in von Barth-Hedin form [\onlinecite{jellium}].
Total and partial densities of states (DOS) together with band dispersions can 
be seen in Fig.~\ref{fig6} for LDA (dashed lines) and LDA$^\prime$ (solid lines).
From top to bottom on Fig.~\ref{fig6} there are MnO, CoO and NiO systems.
As reported earlier for NiO [\onlinecite{cLDADMFT}] LDA$^\prime$ approach changes
charge transfer energy $|E_d-E_p|$,
where $E_d$ and $E_p$ are, roughly speaking, one electron energy positions of
transition metal 3d and O-2p bands. In Fig.~\ref{fig6} the same tendency 
for MnO and CoO oxides is seen. For MnO it increases about 0.5 eV and for CoO 
about 1 eV, similar to NiO. Almost rigid shift of O-2p bands down in energy 
is observed here, while transition metal 3d states remain almost the same near 
the Fermi level.

One should mention that (to our knowledge) transition metal 4s states were 
never included previously into LDA+DMFT calculations for these transition metal 
oxides. Apparently, this happened because they were reasonably
assumed to be weakly correlated and thus projected out from corresponding LDA 
Hamiltonian. However, transition metal 4s states are rather close to the Fermi 
level for LDA bands and getting even closer for LDA$^\prime$ ones.
They can be seen on Fig.~\ref{fig6} as lowest unoccupied states which are 
touching the Fermi level for MnO near $\Gamma$-point and less than 1eV above 
the Fermi level for CoO and NiO.

\subsection{LDA+DMFT and LDA$^\prime$+DMFT spectral functions}

Everywhere in this paper we employ Hirsh-Fye quantum Monte-Carlo algorithm 
[\onlinecite{QMC}] as impurity solver for DMFT equations. To set up DMFT lattice problem 
we use corresponding LDA and LDA$^\prime$ Hamiltonians, which include all states 
(without any projecting, as was done e.g. Ref.~[\onlinecite{gfdc}]).
Inverse temperature was taken $\beta=5$eV$^{-1}$, with 80 time slices
for NiO, while for MnO and CoO we used $\beta=10$eV$^{-1}$ with 120 and 160 
time slices respectively. Monte Carlo sampling was done with 10$^6$ sweeps.
The use of rather high temperatures does not lead to any qualitative effects in 
the results, allowing us to avoid unnecessary computational efforts. 
Parameters of Coulomb interaction were chosen as typical for MnO, CoO and NiO 
[\onlinecite{Karolak,Kunes}]: $U$=8~eV and $J$=1~eV. 
Both FLL(SC) and FLL(LDA) double counting definitions were applied for all 
materials. Respective $E_{dc}$ values are given in Table~\ref{tab1}. 

\begin{figure*}[!ht]
\includegraphics[width=.405\textwidth]{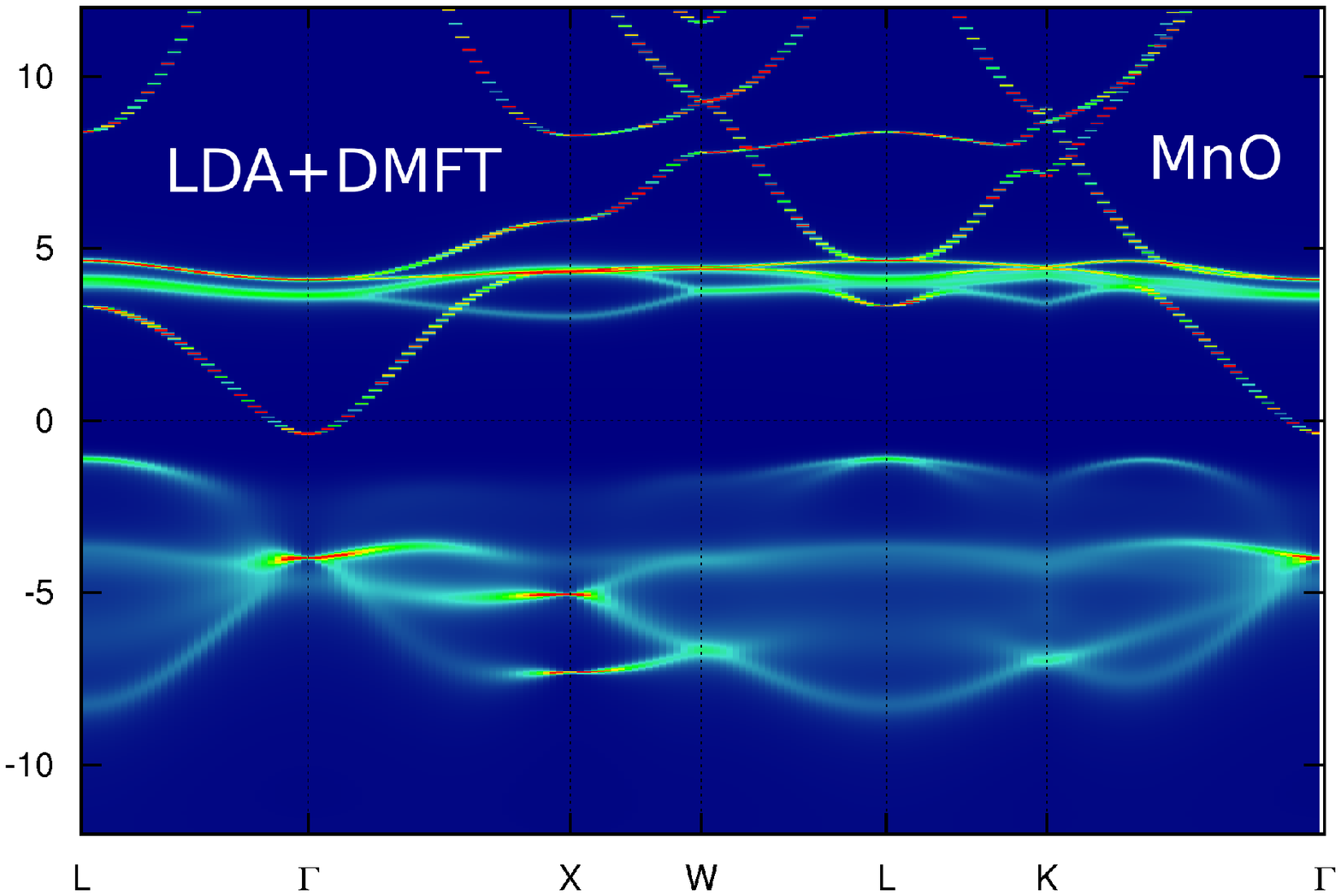}
\includegraphics[width=.45\textwidth]{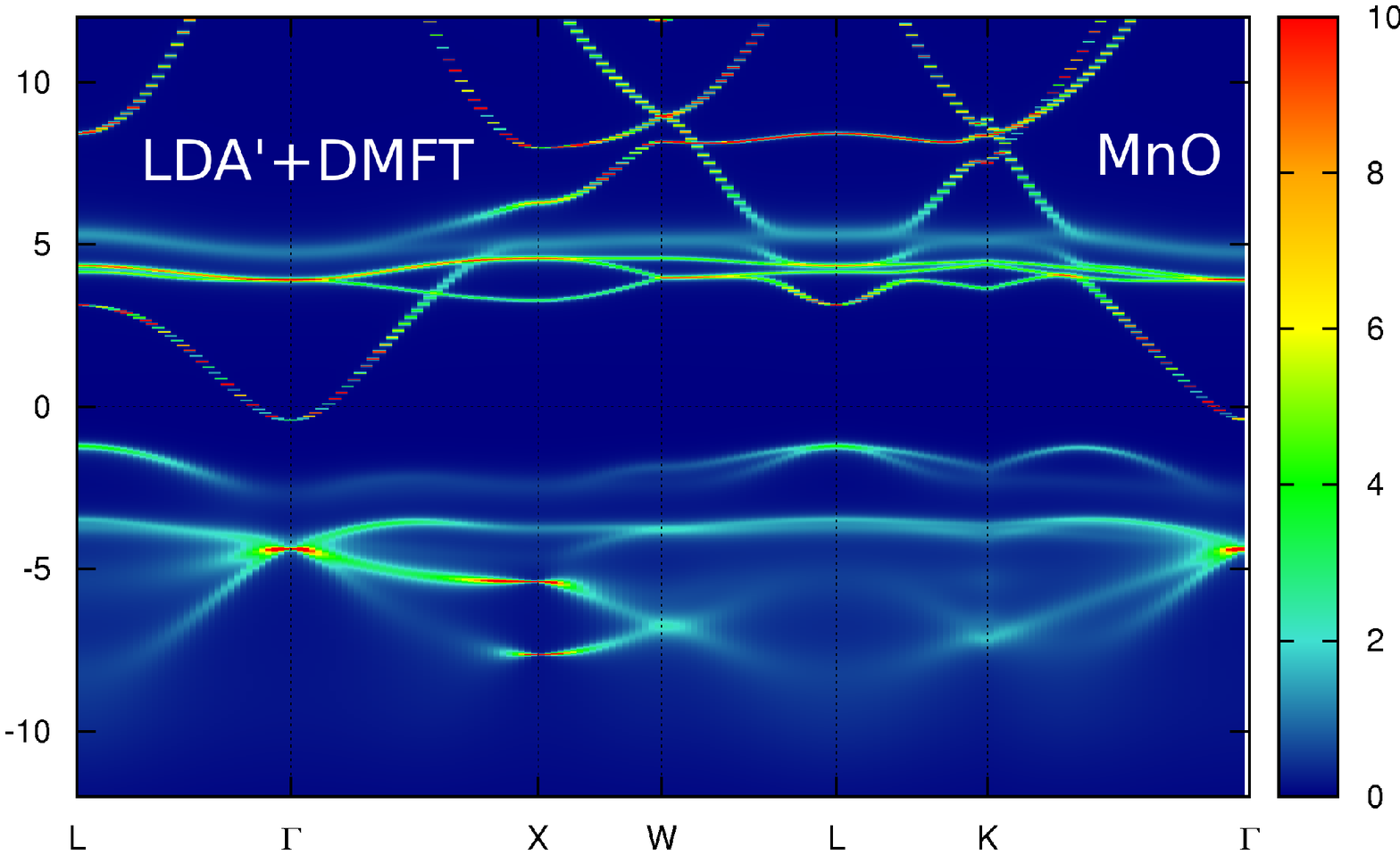}
\includegraphics[width=.405\textwidth]{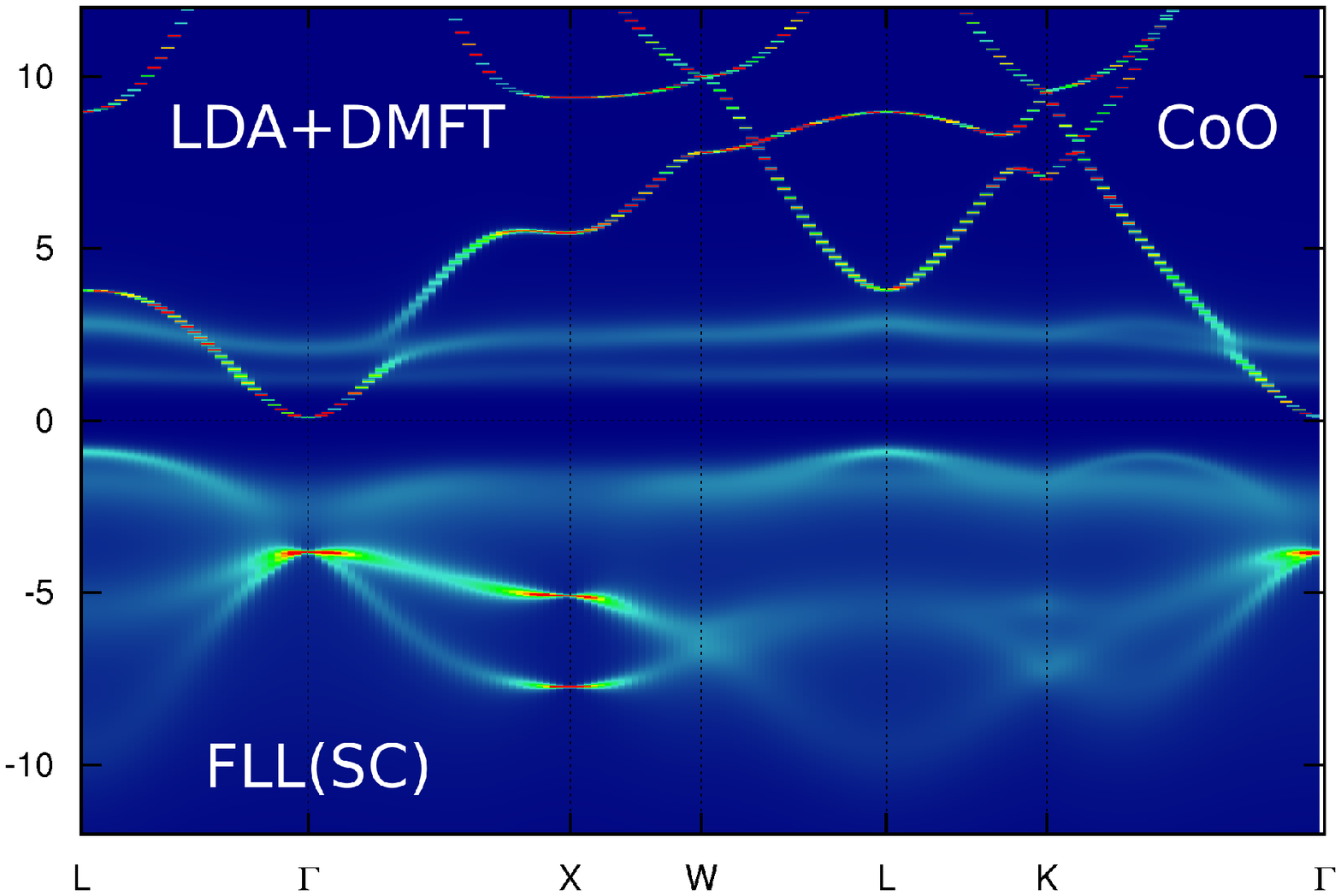}
\includegraphics[width=.45\textwidth]{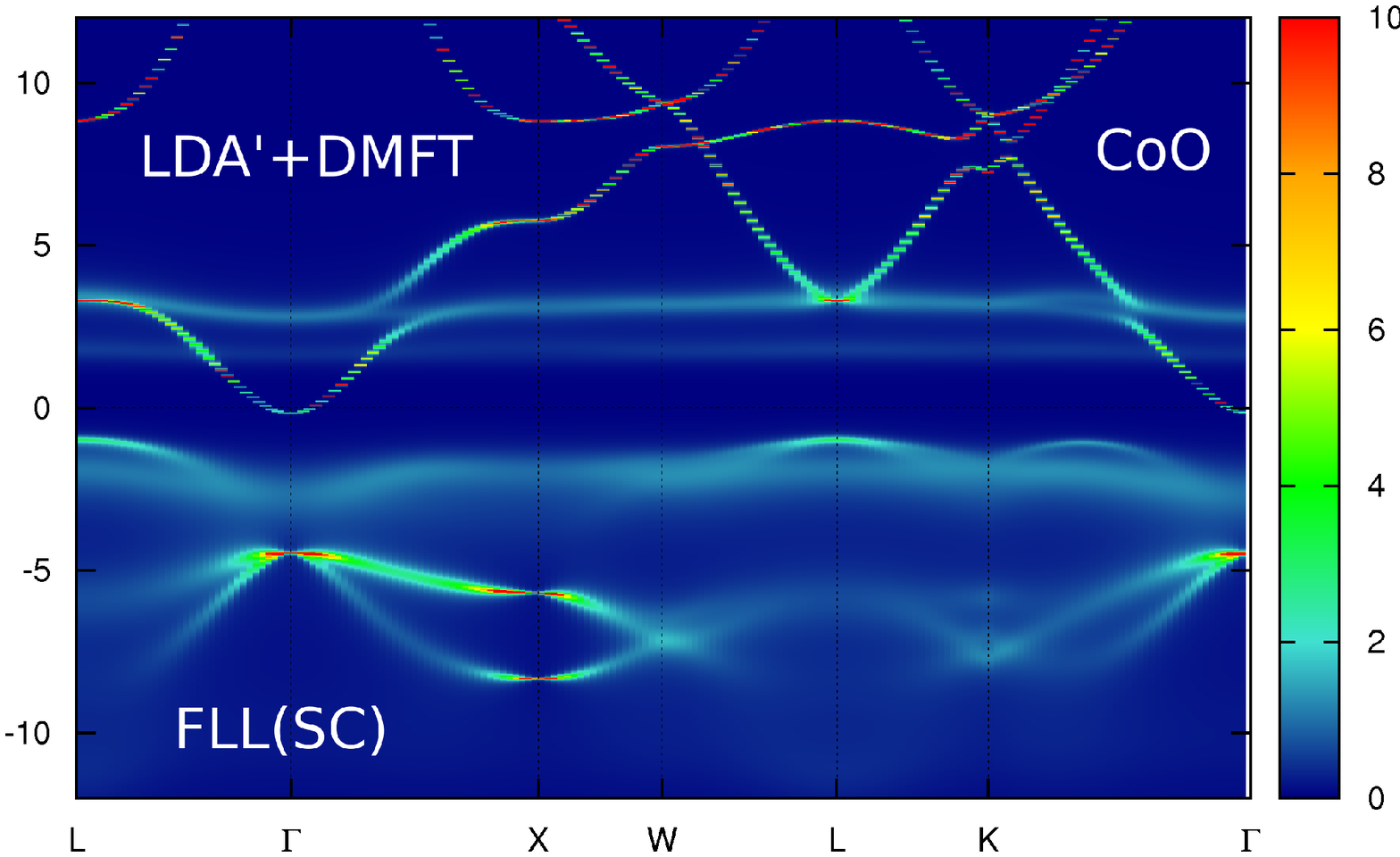}
\includegraphics[width=.405\textwidth]{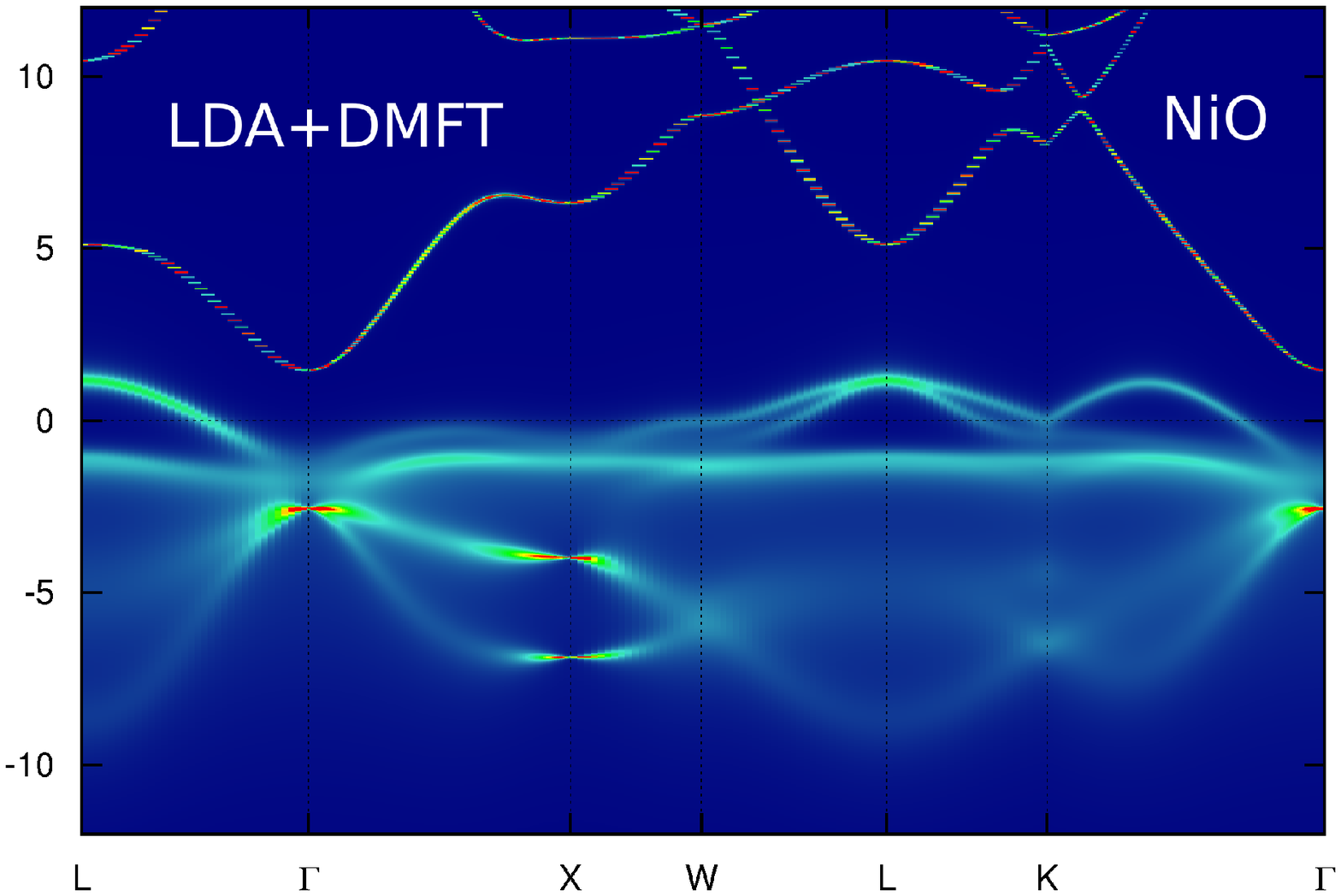}
\includegraphics[width=.45\textwidth]{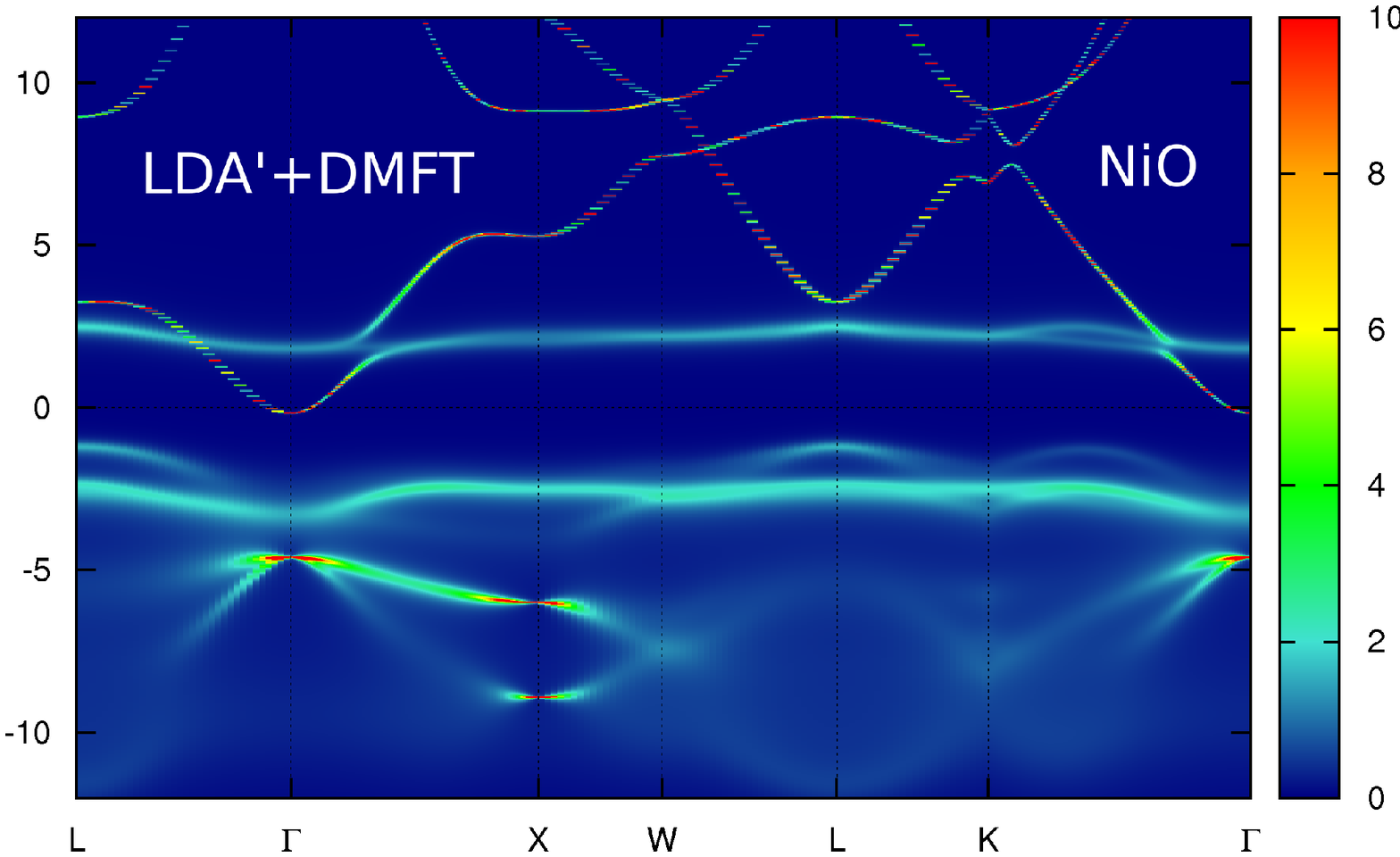}
\caption{(Colour online) Comparison of LDA+DMFT (left column) and 
LDA$^\prime$+DMFT (right column) calculated spectral density functions for 
MnO (upper row), CoO (middle row) and NiO (lower row), with FLL(SC) double
counting correction. Fermi level is zero.
}
\label{fig7}
\end{figure*}

To obtain DMFT(QMC) densities of states (DOS) at real energies, we employed
the maximum entropy method (MEM) [\onlinecite{MEM}]. Then one can get DMFT self-energy 
on the real frequency axis by using Pade approximants for analytical 
continuation. Further on it was checked that ``Pade'' DOS'es are identical 
to ``MEM'' DOS'es. Once $\Sigma(\omega)$ is obtained, one can input it into 
Eq.~(\ref{Dyson_for_DMFT}) and obtain the spectral density function 
$A({\bf k},\omega)=-\frac{1}{\pi}{\rm Im}G({\bf k},\omega)$.
Corresponding maps of spectral density functions, representing effective
band structure of these compounds, are given in Fig.~\ref{fig7}.
Left column of Fig.~\ref{fig7} presents LDA+DMFT results and the right one -- 
LDA$^\prime$+DMFT for MnO (upper panels), CoO (middle panels) and NiO (lower 
panels).

\subsection{LDA+DMFT and LDA$^\prime$+DMFT DOS}

\begin{figure*}[!ht]
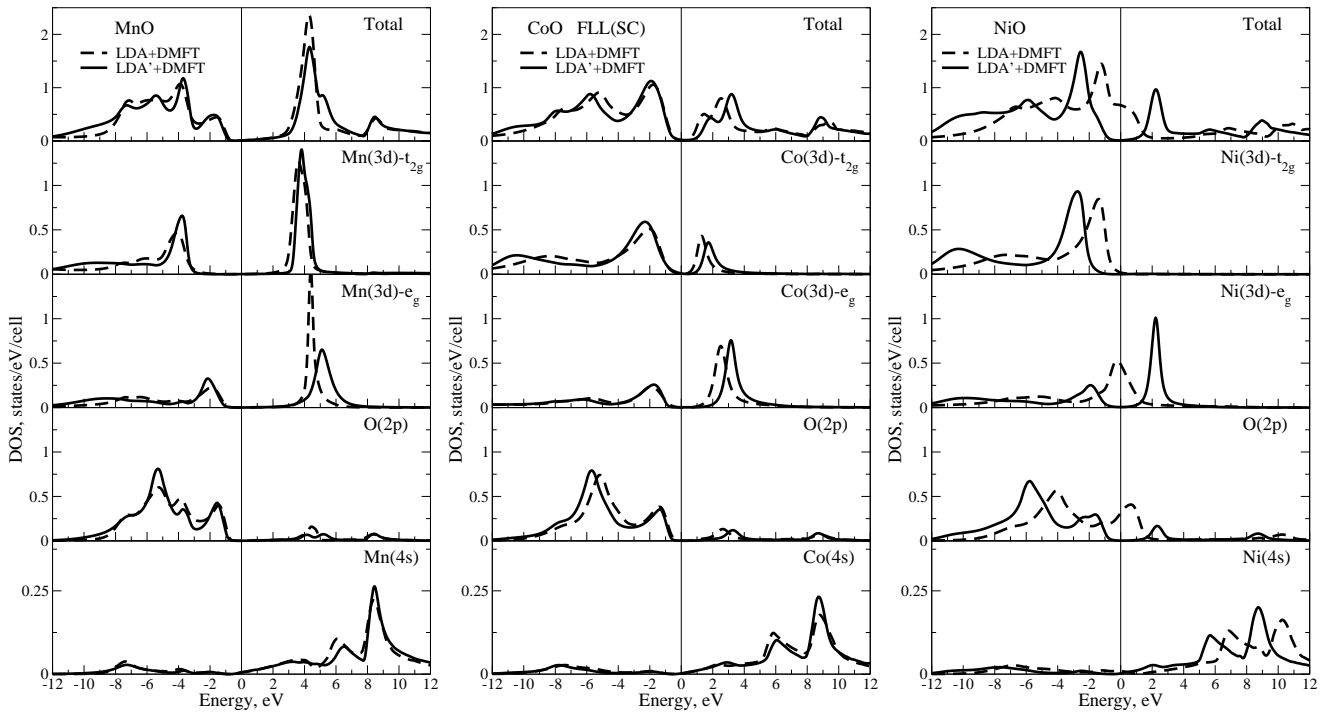

\includegraphics[width=.32\textwidth]{MnO_comp_FLL_SC_pade_DOS_with_O2p_Mn4s_panel.eps}
\includegraphics[width=.32\textwidth]{CoO_comp_FLL_SC_pade_DOS_with_O2p_Co4s_panel.eps}
\includegraphics[width=.32\textwidth]{NiO_comp_FLL_SC_pade_DOS_with_O2p_Ni4s_panel.eps}
\caption{Comparison of LDA+DMFT (dashed lines) and LDA$^\prime$+DMFT (solid lines) 
densities of states for MnO (left panel), CoO (middle panel) and 
NiO (right panel), with FLL(SC) double counting correction. Fermi level is zero.
}
\label{fig8}
\end{figure*}

In Fig.~\ref{fig8} we present densities of states obtained by LDA+DMFT 
(dashed lines) and LDA$^\prime$+DMFT (solid lines). 
The left panel corresponds to MnO, middle one to CoO and the left one to NiO. 
Upper row shows total densities of states, while in other rows
we show the contributions of the most important electron
states -- $t_{2g}$ and $e_g$ subshells for 3d transition metal, oxygen 2p states 
and also transition metal 4s states.
\begin{figure}[!hb]
\includegraphics[width=.32\textwidth]{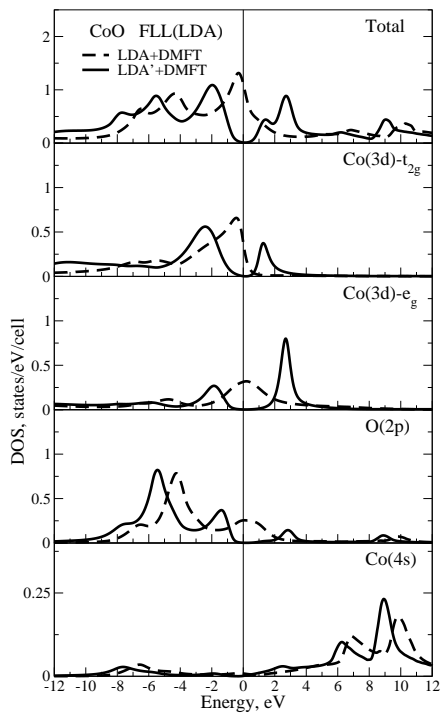}
\caption{Comparison of LDA+DMFT (dashed lines) and LDA$^\prime$+DMFT 
(solid lines) calculated densities of states for
CoO with FLL(LDA) double counting correction. Fermi level is zero.
}
\label{fig9}
\end{figure}

First we focus on MnO case which is perhaps the simplest one among these three.
The O-2p states are located between -9 and -4 eV (see 
Figs.~\ref{fig7}~and~\ref{fig8}). 
Then comes lower Hubbard band, which consists of Mn-3d $t_{2g}$ and $e_g$
contributions at -4eV and -2.3 eV correspondingly.
On spectral function maps LHB is rather wide non dispersive band
at these energies. Then we see the so called Zhang-Race band -- the bound
state which appears when strongly interacting band is hybridized with charge 
reservoir.
This band can be seen as a peak at -1.5 eV in O-2p states together with 
Mn-3d $e_g$ states. Then, between the Zhang-Race band and the upper Hubbard band
there is a gap for Mn-3d states about 3.5 eV for both
LDA+DMFT and LDA$^\prime$+DMFT, which agrees pretty well with experimental 
spectra (see below). UHB is located above 4 eV and, where $t_{2g}$ and $e_g$ 
contributions can not be separated in energy.

Spectral density map of Fig.~\ref{fig7} (upper row) show some rather well 
defined band of MnO, which touches the Fermi level in the $\Gamma$-point.
This band is nothing else but Mn-4s. It is seen from Fig.~\ref{fig8}, that 
most of the Mn-4s spectral weight is actually well above 5 eV. Below there is 
some rather low intensity tail, which goes through the gap between the upper 
Hubbard band and the Zhang-Rice band.
Its intensity is at least one order of magnitude lower, than intensities of 
other contributions to DOS. 

Consider next CoO (middle row of Fig.~\ref{fig7} and middle panel of
Fig.~\ref{fig8}). We see that both LDA+DMFT and LDA$^\prime$+DMFT results are 
quite similar. There is some difference in the UHB, where Co-3d $t_{2g}$ and 
$e_g$ contributions can now be separated and in Fig.~\ref{fig8} two almost 
nondispersive bands around 2 and 3 eV above the Fermi level are clearly seen.
The gap between  Zhang-Rice band and UHB is about 0.5 eV larger (about 4 eV)
for LDA$^\prime$+DMFT results. 

One should note, that LDA+DMFT calculation with FLL(LDA) double counting 
produces the metallic solution for CoO, as seen from Fig.~\ref{fig9}, which 
qualitatively contradicts the experiments. On the contrary, LDA$^\prime$+DMFT 
gives the correct insulating state.

Note, that both in CoO and NiO the behavior of 4s bands is similar to
that discussed above for the case of MnO. Spectral density maps of Fig.~\ref{fig7}
show the presence of these bands within the charge transfer gap, though the
partial density of states due to these bands within the gap is almost 
negligible (cf. Fig.~\ref{fig8}).

To sum up, we stress that both MnO and CoO  within LDA$^\prime$+DMFT are
consistently demonstrated to be charge transfer insulators (in contrast to the
conventional LDA+DMFT in the case of CoO). The similar behavior was 
obtained earlier for NiO in Ref. [\onlinecite{cLDADMFT}].
Here we presented more complete LDA$^\prime$+DMFT results for NiO, with
both FLL(LDA) and FLL(SC) double counting corrections. Conventional
LDA+DMFT calculations predict NiO to be metallic in contrast to experiment,
while LDA$^\prime$+DMFT gives charge transfer insulating solution for NiO for 
both FLL(LDA) and FLL(SC) double counting correction.
All other features of NiO LDA$^\prime$+DMFT band structure are quite similar
to MnO and CoO compounds described above.

\begin{figure*}[!ht]
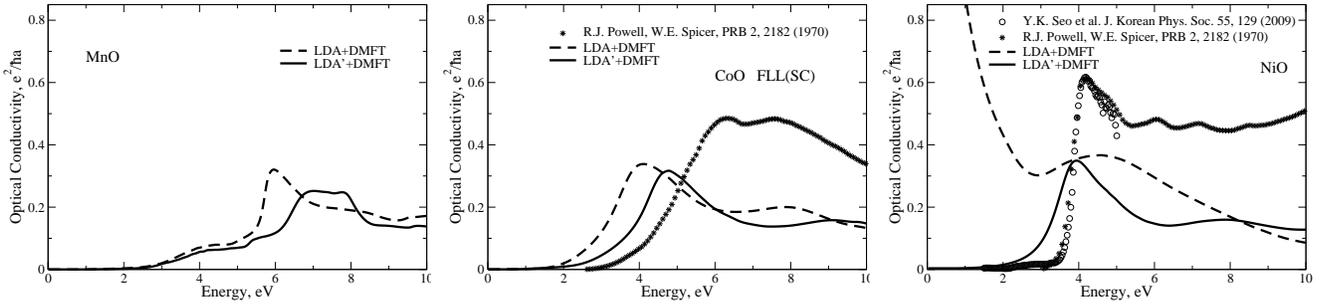

\includegraphics[width=.32\textwidth]{MnO_comp_FLL_SC_Optical_conductivity_Intra_3.eps}
\includegraphics[width=.32\textwidth]{CoO_comp_FLL_SC_Optical_conductivity_Intra_Exp.eps}
\includegraphics[width=.32\textwidth]{NiO_comp_FLL_SC_Optical_conductivity_Intra_Exp_5a.eps}
\caption{Comparison of experimental (circles, stars) and calculated  
LDA+DMFT (dashed lines) and LDA$^\prime$+DMFT (solid lines) optical 
conductivities for MnO (left panel), CoO (middle panel) and NiO (right panel).}
\label{fig10}
\end{figure*}

\subsection{LDA+DMFT and LDA$^\prime$+DMFT optical conductivity}

Metallic or insulating behavior can be explicitly demonstrated by calculations
of optical conductivity. Below we present our results for optical conductivity
behavior of MnO, CoO and NiO within LDA+DMFT and LDA$^\prime$+DMFT, allowing
us also to analyze the influence of transition metal 4s states
on dielectric properties of these oxides. In our calculations we used the
following expression for optical conductivity, valid in DMFT [\onlinecite{AdvPr}]:
\begin{eqnarray}
\sigma_{xx}(\omega)&=&\frac{\pi e^2}{2\hbar a} \int\limits^{\infty}_{\infty} d\varepsilon \frac{f(\varepsilon)-f(\varepsilon-\omega)}{\omega}\times\nonumber\\
&\frac{1}{N}& \sum\limits_{ij\vec k \sigma} 
\left(\frac{\partial \varepsilon^i_{\vec k}}{\partial k_{x}}\right)\left(\frac{\partial \varepsilon^j_{\vec k}}{\partial k_{x}}\right)
A^{ij}_{\vec k}(\varepsilon)A^{ji}_{\vec k}(\varepsilon-\omega).
\end{eqnarray}
Here $e$ is electron charge, $a$ is the lattice constant of corresponding 
compound, $f(\varepsilon)$ -- Fermi function, $\varepsilon_{\vec k}$ -- 
band dispersion, \modified{$A^{ij}_{\vec k}(\varepsilon)$ corresponding (LDA+DMFT or 
LDA$^\prime$+DMFT) spectral density function matrix ($i,j$ are the band indices).
During our calculations we found that main contribution to optical conductivity
is due to intra-orbital optical transitions. Inter-orbital optical transitions give
less than 5\% of optical conductivity intensity in frequency range
used in our calculations.
Also in the present work we neglect possible effects due to optical matrix elements.
}
Calculated theoretical curves obtained in conventional LDA+DMFT (dashed line) 
and within LDA$^\prime$+DMFT (solid line) are presented in Fig.~\ref{fig10} 
for MnO (left panel), CoO (middle panel) and NiO (right panel).

From Fig.~\ref{fig10} we see, that within LDA$^\prime$+DMFT (solid line) all 
materials are insulators. Despite the presence of transition metal 4s states 
close to the Fermi level, possible Drude peak due to these states is not 
observed. Conventional LDA+DMFT optical conductivity for NiO shows typical 
metallic behavior, as discussed earlier in the context of DOS behavior.

Now we compare our theoretical results with available experimental data
(with an exception of MnO, where we are not aware of any experimental
results) [\onlinecite{Powell,Seo}]. In Ref.~[\onlinecite{Powell}] only
experimental data for optical constants $n(\omega)$ and $k(\omega)$ were 
presented. The optical conductivity in units of $\frac{e^2}{\hbar a}$ 
(which is about 5.8$\times 10^{3}\Omega^{-1}cm^{-1}$ for given monooxides) can
be recalculated from these data using as 
$\sigma(\omega)=\frac{n(\omega)k(\omega)}{2\pi}\omega\alpha^{-1}\frac{a}{c}$,
where $\alpha$ is fine structure constant, $a$ -- lattice constant and $c$ -- 
speed of light. Corresponding curves are shown in Fig.~\ref{fig10} by stars. 
For NiO there are more recent experimental data of Ref.~[\onlinecite{Seo}],
shown with circles. 
One observes that below the leading absorption edge for CoO and NiO there exist
rather long absorption tails with low intensity. We associate these tails with 
contribution of Co and Ni 4s states. For NiO the overall agreement of
LDA$^\prime$+DMFT results with experimental data is quite satisfactory. 
For CoO theoretical absorption edge is about 1 eV lower than experimental one. 
However, this can probably 
be corrected introducing the larger value of Coulomb interaction $U$. Recent
constrained RPA study produced it to be 10.8 eV [\onlinecite{sinfty}], in 
contrast to 8 eV used in our calculations.

\begin{figure}[!ht]
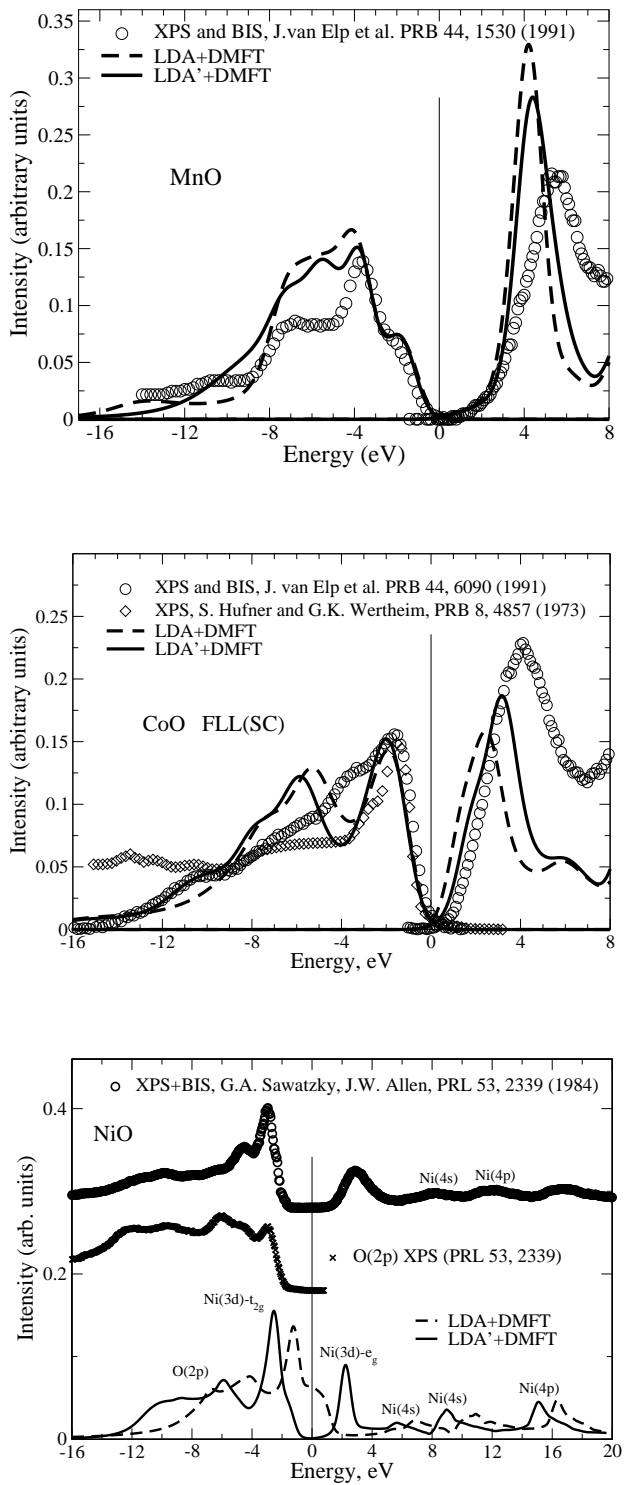

\includegraphics[width=.45\textwidth]{MnO_comp_FLL_SC_exp_5_black2.eps}
\newline

\vspace{.7cm}
\includegraphics[width=.45\textwidth]{CoO_comp_FLL_SC_exp_black.eps}\newline

\vspace{.7cm}
\includegraphics[width=.45\textwidth]{NiO_DMFT_comp_FLL_comp_exp_SA_2.eps}

\caption{Comparison of LDA+DMFT (dashed lines) and LDA$^\prime$+DMFT 
(solid lines) spectra with XPS and BIS
experimental data (circles, diamonds, crosses)
for MnO (upper panel), CoO (middle panel) and NiO (lower panel). Fermi level 
is zero.
}
\label{fig11}
\end{figure}

\subsection{Comparison of LDA+DMFT and LDA$^\prime$+DMFT results with 
X-ray experiments}

Now we compare our results for DOS with XPS and BIS experiments of Refs.
[\onlinecite{MnO_XPS_BIS,CoO_XPS,CoO_XPS_BIS,NiO_XPS_BIS}].
In Fig.~\ref{fig11} LDA+DMFT (dashed lines) and LDA$^\prime$+DMFT 
(solid lines) valence and conduction bands spectra are directly compared
with spectra for MnO (upper panel), CoO (middle panel) and NiO (lower panel).
Theoretical spectra were obtained by multiplication of DOS by Fermi distribution
and Gaussian broadening with experimental temperature and resolution. 

General structure of spectra is similar for all three compounds.
From -14 to -4 eV there are O-2p states, then comes lower Hubbard band at 
about -3 eV. On the high energy slope of the LHB there we can see a 
shoulder-like structure, which is nothing else but Zhang-Rice band.
Around the Fermi level there is insulating gap. The size of the gap is very 
well reproduced for MnO by both LDA+DMFT and LDA$^\prime$+DMFT.
For CoO it looks like $U$ value chosen is a bit too small (as discussed earlier),
however LDA$^\prime$+DMFT spectra gives gap size closer to the experiment.
For NiO conventional LDA+DMFT gives metallic solution, while LDA$^\prime$+DMFT 
produces CTI solution with correct energy gap size.
Experimental positions of the upper Hubbard bands are rather well described
by LDA$^\prime$+DMFT.
Since experimental data for NiO goes far above
the Fermi level one can identify these high energy structures
as Ni-4s and Ni-4p states contributions.

On Fig.~\ref{fig11} one can see that experimental conduction band low energy 
threshold has a rather long low intensity tail which goes down to the Fermi level.  
Because of that there is some asymmetry of the gap. We suggest
that this asymmetry of the gap originates from transition metal 
4s states, which touch the Fermi level from above, as described earlier.

\section{Strongly correlated metals}
\label{scm}

\subsection{LDA and LDA$^\prime$ band structure}

\begin{figure*}[!ht]
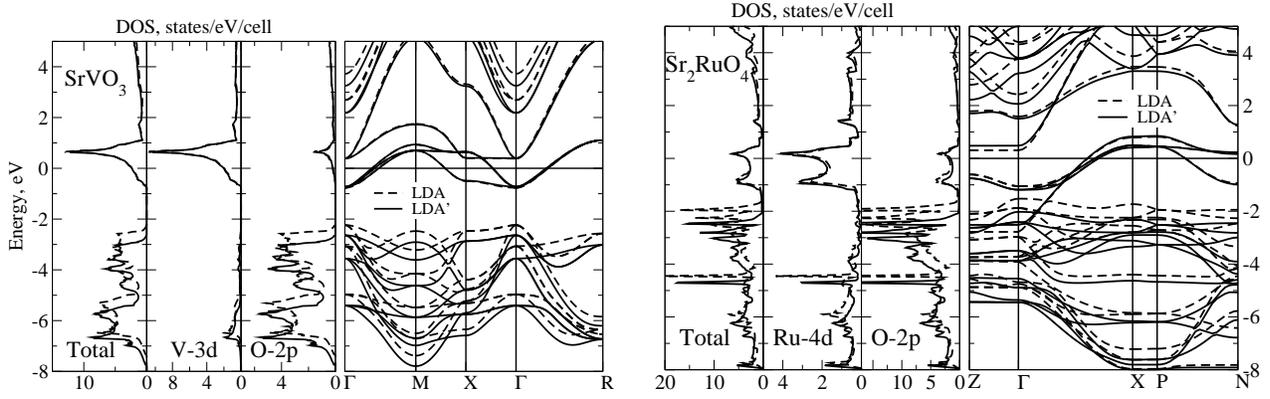

\includegraphics[width=.45\textwidth]{SrVO3_LDA_bands_and_DOS_comp.eps}
\hspace{0.3cm}
\includegraphics[width=.45\textwidth]{Sr2RuO4_LDA_bands_and_DOS_comp.eps}
\caption{LDA (dashed lines) and LDA$^\prime$ (solid lines) band dispersions 
for SrVO$_3$ (left panel) and Sr$_2$RuO$_4$ (right panel). Fermi level is zero.
}
\label{fig1}
\end{figure*}

Strontium vanadate SrVO$_3$ is perhaps one of the most simple
paramagnetic strongly correlated metallic systems.
There is no wonder that it is widely used as a test system for various
LDA+DMFT based numerical techniques [\onlinecite{srvo1,srvo2,srvo3,srvo4}].  
SrVO$_3$ has ideal cubic perovskite structure with one d-electron in V-3d shell
within triply degenerated $t_{2g}$ subshell.
LDA and LDA$^\prime$ band structure calculations are performed as described in 
Refs.~[\onlinecite{srvo1,srvo2,srvo3,srvo4}]
via LMTO method with von Barth-Hedin exchange correlation energy 
[\onlinecite{jellium}].

\begin{figure}[!ht]
\includegraphics[width=.45\textwidth]{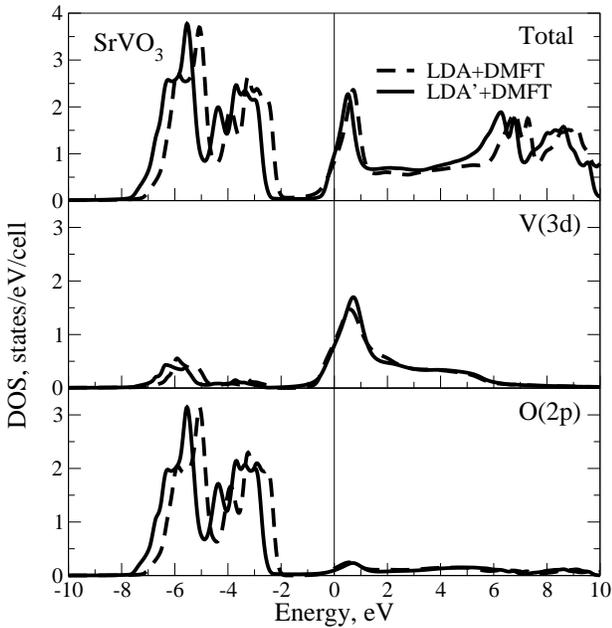}
\caption{Densities of states calculated with LDA+DMFT (dashed lines) and 
LDA$^\prime$+DMFT (solid lines) for SrVO$_3$: V-3d states - upper panel,
O-2p states - lower panel. Fermi level is zero.
}
\label{fig2}
\end{figure}

The 3d bands of vanadium cross the Fermi level, while oxygen 2p states are 
at -8 -- -2 eV i.e. much lower than the Fermi level (see Fig.~\ref{fig1}, 
left panel, dashed lines). If we exclude  $E_{\rm xc}^{LDA}$ contribution 
for V-3d states as described in Sec.~III, we obtain the
LDA$^\prime$ band structure shown in Fig.~\ref{fig1} (left panel, solid lines).
Similar to Ref.~[\onlinecite{cLDADMFT}] within LDA$^\prime$ approach energy 
splitting between V-3d and O-2p bands $|E_d-E_p|$ becomes larger, than in 
conventional LDA. Since the total number of electrons is fixed,
the increase of $|E_d-E_p|$ LDA$^\prime$ is related to O-2p bands going down in 
energy by about 0.5 eV, with V-3d states remaining almost unchanged.
One should mention here also, that the overall bandshapes are practically not 
changed in comparison with the conventional LDA bands. The same is true of 
course for densities of states presented on the left panel of Fig.~\ref{fig1}.

Another example of paramagnetic strongly correlated metallic system widely
treated by LDA+DMFT is Sr$_2$RuO$_4$ with Ru-4d$^4$ $t_{2g}$ subshell
(see Ref.~[\onlinecite{srruo1}] and references therein). 
Sr$_2$RuO$_4$ is a layered perovskite with an ideal body-centered tetragonal 
crystal structure. For LDA and LDA$^\prime$ calculations we used settings 
described in Ref.~[\onlinecite{srruo1}]. LDA (dashed lines) and LDA$^\prime$ 
(solid lines) band dispersions and DOS'es are plotted in  Fig.~\ref{fig1} 
(right panel). The picture here is not that simple as for SrVO$_3$.
The Ru-4d states, crossing the Fermi level, almost preserve their energy 
positions and dispersions within LDA$^\prime$. However LDA$^\prime$ leads to
$|E_d-E_p|$ splitting, because of non-uniform narrowing of O1-2p and O2-2p 
states, together with the slight shift of O2-2p states. In total $|E_d-E_p|$ 
energy splitting is about 0.5 eV larger for LDA$^\prime$ than in conventional
LDA.

\begin{figure}[!ht]
\includegraphics[width=.45\textwidth]{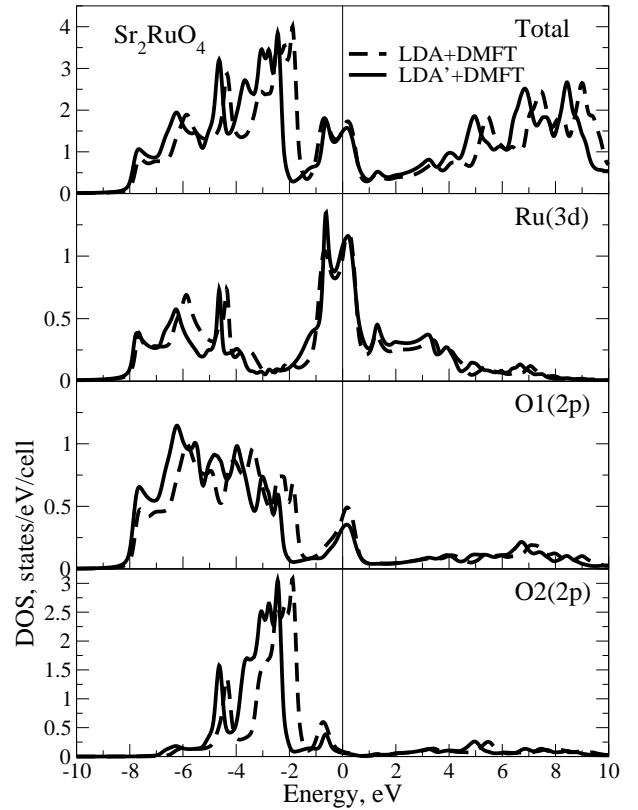}
\caption{Densities of states calculated with LDA+DMFT (dashed lines) and LDA$^\prime$+DMFT (solid lines)
for Sr$_2$RuO$_4$: Ru-4d states - upper panel,
O-2p states - middle and lower panels. Fermi level is zero.
}
\label{fig3}
\end{figure}

\subsection{LDA+DMFT and LDA$^\prime$+DMFT DOS}

In contrast to previous works (Refs.~[\onlinecite{srvo1,srvo2,srvo3,srvo4,srruo1}])
we used here the full TB-LMTO-ASA calculated LDA and LDA$^\prime$ Hamiltonians,
employing none of the widely used projection techniques. In QMC calculations 
inverse temperature was taken to be $\beta=10$eV$^{-1}$, with 80 time slices
for SrVO$_3$, while for Sr$_2$RuO$_4$ we used $\beta=15$eV$^{-1}$, with 64 time 
slices. Coulomb parameters were taken to be $U$=6.0 eV and $J$=0.7 eV 
[\onlinecite{gfdc}] for  SrVO$_3$ and 3.2 eV and 0.7 eV for Sr$_2$RuO$_4$ 
respectively [\onlinecite{srruo1}]. Number of Monte Carlo sweeps was of the 
order of 10$^6$. To obtain DMFT(QMC) [\onlinecite{QMC}] densities 
of states at real energies, we again employed the maximum entropy method 
[\onlinecite{MEM}]. To get corresponding DMFT O-2p densities of states the
method of Pade approximants was applied to do analytic continuation for 
DMFT self-energy from Matsubara to real frequencies, with further crosschecking
of ``MEM'' and ``Pade'' DOS'es to ensure the quality of restored self-energy 
for real frequencies.

In Figs.~\ref{fig2} and~\ref{fig3} we present the total and partial densities 
of states for SrVO$_3$ and Sr$_2$RuO$_4$ calculated by the conventional
LDA+DMFT (dashed lines) and LDA$^\prime$+DMFT (solid lines).
For both systems LDA$^\prime$+DMFT results show
lower positions of O-2p states in comparison with LDA+DMFT.
However, for Sr$_2$RuO$_4$ it does not reduce just to a rigid shift of oxygen
states by about 0.5 eV, as in the case of SrVO$_3$, but is the combination of 
some small shift with non-uniform narrowing of oxygen bands.
Thus, for Sr$_2$RuO$_4$ only the high energy threshold of O-2p
states moves down by 0.5 eV.

As opposed to Refs.~[\onlinecite{srvo1,srvo2,srvo3,srvo4}]
in both calculations for SrVO$_3$ we observe very smooth
upper and lower Hubbard bands in V-3d DOS (upper panel of Fig.~\ref{fig2}).
This agrees well with the full orbital calculations reported in 
Ref.~[\onlinecite{gfdc}]. Also in Ref.~[\onlinecite{gfdc}] it is shown that 
smaller value of $E_{dc}$ (if $E_{dc}$ is treated as free parameter)
moves oxygen states down in energy, which leads to better agreement with 
experiment (see the next paragraph).

\subsection{Comparison of LDA+DMFT and LDA$^\prime$+DMFT results 
with X-ray experiments}

\begin{figure}[!ht]
\includegraphics[width=.45\textwidth]{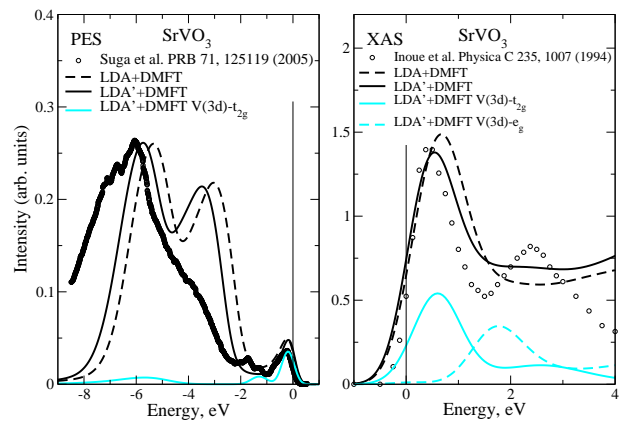}
\caption{(Colour online) Comparison of LDA+DMFT (dashed lines) and LDA$^\prime$+DMFT (solid lines) calculated spectra
for SrVO$_3$ with experimental data. With cyan colour LDA$^\prime$+DMFT
$t_{2g}$ (solid line) and $e_g$ (dot-dash line) V-3d contributions are shown.
 Fermi level is zero.}
\label{fig4}
\end{figure}

\begin{figure}[!ht]
\includegraphics[width=.45\textwidth]{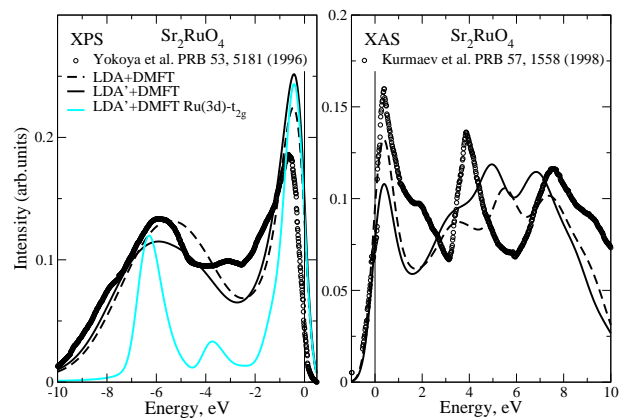}
\caption{(Colour online) Comparison of LDA+DMFT (dashed lines) and LDA$^\prime$+DMFT (solid lines) calculated spectra
for Sr$_2$RuO$_4$ with experimental data.
With cyan colour LDA$^\prime$+DMFT $t_{2g}$ Ru-4d contribution is shown.
Fermi level is zero.}
\label{fig5}
\end{figure}

In Figs.~\ref{fig4} and~\ref{fig5} LDA+DMFT (dashed lines) and LDA$^\prime$+DMFT (solid lines)
calculated spectra for SrVO$_3$ and Sr$_2$RuO$_4$ correspondingly are drawn.
To get theoretical spectra from total DOS Gaussian broadening to simulate experimental resolution
and Lorentzian broadening to simulate lifetime effects together with multiplication
with Fermi distribution function were performed as described elsewhere 
[\onlinecite{srvo1,srvo2,srvo3,srvo4,srruo1}].
On the figures emission (left side) and absorption (right side) spectra are plotted.

For both systems we have reasonable agreement with experimental data (circles) for valence and
conducting bands [\onlinecite{srvo2,SrVO3_XAS,Sr2RuO4_XPS,Sr2RuO4_XAS}] (see Figs.~\ref{fig4}, ~\ref{fig5}).
However strength of quasiparticle peak is a bit overestimated
for valence band and underestimated for conduction band in both LDA+DMFT and LDA$^\prime$+DMFT methods.
The LDA$^\prime$+DMFT results give slightly better energy position of
O-2p states in comparison to LDA+DMFT. In general obtained by LDA$^\prime$+DMFT results
are in agreement with previous LDA+DMFT works (see Refs.~[\onlinecite{srvo1,srvo2,srvo3,srvo4,srruo1}]).

To demonstrate presence of well known lower Hubbard band at -1.5 eV for  SrVO$_3$ 
[\onlinecite{srvo1,srvo2,srvo3,srvo4}]
on left panel of Fig.~\ref{fig4} V-3d $t_{2g}$ contribution is shown by cyan line.
In Fig.~\ref{fig4} (right panel) for SrVO$_3$ instead of upper Hubbard band around 2.5 eV
LDA$^\prime$+DMFT shows rather broad shoulder. This shoulder is formed
by $t_{2g}$ (solid cyan line) and $e_g$ (dot-dash cyan line) V-3d contributions
which corresponds to previous works [\onlinecite{srvo1,srvo2,srvo3,srvo4}].
However the $e_g$ subband in our case is also modified by correlations. It is shifted up on about 1 eV
(as should be for completely empty states) and it has smaller width compared to the LDA one.
For Sr$_2$RuO$_4$ it is known that correlations lead to formation of
lower Hubbard band satellite near -3 eV [\onlinecite{srruo1}]. This satellite is
also seen in the  LDA$^\prime$+DMFT results on the right panel of Fig.~\ref{fig5}
and is formed essentially by Ru-4d $t_{2g}$ states (cyan line).

\begin{table*}[!hb]
\caption{LDA and LDA$^\prime$ occupancies and corresponding values of LDA+DMFT and LDA$^\prime$+DMFT
double counting terms (eV) for systems under consideration.}
\begin{ruledtabular}
\begin{tabular}{|c|c|c|c|c|c|c|}
 Compound               & n$_{LDA}$ & n$_{LDA'}$ & LDA+DMFT & LDA+DMFT & LDA'+DMFT & LDA'+DMFT       \\
                        &           &            & FLL(LDA) & FLL(SC)   & FLL(LDA) & FLL(SC)         \\
\hline
SrVO$_3$                &  2.61     & 2.44       & 12.33    & 11.99     & 10.35    & 10.92           \\
\hline
Sr$_2$RuO$_4$           &  5.65     & 5.39       & 14.32    & 14.60     & 12.92    & 13.73           \\
\hline
MnO                     &  5.59     & 5.43       & 39.05    & 35.49     & 36.62    & 35.30           \\
\hline
CoO                     &  7.60     & 7.41       & 54.28    & 50.90     & 51.42    & 50.49           \\
\hline
NiO                     &  8.54     & 8.34       & 60.90    & 62.01     & 57.91    & 58.13           \\
\end{tabular}
\end{ruledtabular}
\label{tab1}
\end{table*}

\section{5. Conclusion}
\label{concl}

This work continues our research of the double counting problem
arising within the LDA+DMFT computational scheme.
The problem appears because some portion of local
electron-electron interaction is already present in LDA calculations.
Since DMFT gives exact local solution of the Hubbard-like model
one should avoid double counting between LDA and DMFT
local electronic interactions.
Despite 15 years of developing of the LDA+DMFT method still there are 
no unique definition of this double counting term.
This happens because LDA contribution to exchange correlation energy has no diagrammatic expression.
Several different \modified{$ad~hoc$} definitions which are available now work well only in some particular
cases, for some particular compounds.
Sometimes one can get even qualitatively wrong LDA+DMFT solution 
if double counting term is chosen not careful enough.
To \modified{overcome} this problem we proposed consistent LDA$^\prime$+DMFT approach 
[\onlinecite{cLDADMFT}].
It uses natural assumption of explicit exclusion of LDA exchange correlation potential
for correlated electronic shells since anyhow exchange-correlation effects will be accounted later by DMFT.
\modified{Then local interactions left out for correlated states in the LDA$^\prime$ are only Hartree ones.}
After that corresponding double counting term of the LDA$^\prime$+DMFT Hamiltonian
\modified{consistently} must be taken in the local Hartree form (FLL form).

With this paper we present extensive LDA$^\prime$+DMFT investigation of typical representatives
of two wide classes of strongly correlated systems in the paramagnetic phase:
strongly correlated metals (SrVO$_3$ and Sr$_2$RuO$_4$) and charge transfer insulators (MnO, CoO and NiO).
For strongly correlated metals where double counting is not that severe
LDA$^\prime$+DMFT agrees well with traditional LDA+DMFT results with FLL double counting type.
LDA$^\prime$+DMFT gives slightly better position of O-2p states in comparison with experiment.
LDA$^\prime$+DMFT results for charge transfer insulators MnO, CoO and NiO are more interesting.
CoO and NiO systems are found to be metals within conventional LDA+DMFT calculations
while LDA$^\prime$+DMFT gives proper insulating solution. 
Transition metal 4s-states missed in previous LDA+DMFT works on these monooxides
are found to be responsible for charge gap asymmetry around the Fermi level.

Finally one can conclude that proposed by us consistent LDA$^\prime$+DMFT
method works well for both metallic and insulating systems. We believe
that our \modified{LDA$^\prime$+DMFT provides reasonable parameter free treatment of the} double counting problem.

\section{Acknowledgements}

We thank A.I. Poteryaev for providing us QMC code and many helpful discussions.
We are grateful to E.Z. Kuchinskii for providing us more insight into 
calculations of optical conductivity.
This work is partly supported by RFBR grant 11-02-00147 and was performed
within the framework of programs of fundamental research of the Russian 
Academy of Sciences (RAS) ``Quantum mesoscopic and disordered structures'' 
(12-$\Pi$-2-1002) and of the Physics Division of RAS  ``Strongly correlated 
electrons in solids and structures'' (012-T-2-1001).
NSP acknowledges the support of the Dynasty Foundation and International Center 
of Fundamental Physics in Moscow.

\end{document}